\documentclass[letterpaper,english,reprint,aps,notitlepage,superscriptaddress,twocolumn,pra]{revtex4-2}

\usepackage[T1]{fontenc}
\usepackage[utf8]{inputenc}
\usepackage{amsmath}
\usepackage{amssymb}
\usepackage[dvipsnames]{xcolor}
\usepackage{nicefrac}

\colorlet{mylinkcolor}{RoyalPurple}
\colorlet{mycitecolor}{RoyalPurple}
\colorlet{myurlcolor}{RoyalPurple}

\usepackage{hyperref}
\hypersetup{
    linkcolor  = mylinkcolor,
    citecolor  = mycitecolor,
    urlcolor   = myurlcolor,
    colorlinks = true,
    breaklinks = true
}

\makeatletter

\pdfpageheight\paperheight
\pdfpagewidth\paperwidth

\makeatother

\usepackage{babel}

\usepackage{lipsum} 
\usepackage{physics}
\usepackage{mathtools}
\usepackage{stackrel}
\usepackage{amssymb}
\usepackage{bm}
\usepackage{ upgreek }
\usepackage{titlesec}
\usepackage{graphicx}
\usepackage{hyperref}
\usepackage{dsfont}
\usepackage{mathrsfs}
\usepackage[normalem]{ulem}

\newcommand{\rhostd}{\boldsymbol{\rho}}

\newcommand{\superL}{\mathcal{L}}
\newcommand{\superR}{\mathcal{R}}
\newcommand{\superU}{\mathcal{U}}
\newcommand{\rep}{\mathcal{G}}

\newcommand{\cyclic}{\tilde{\rhostd}}
\newcommand{\dbar}{\text{\it{\dj}}}
\newcommand{\unit}{\cdot 2\pi~\text{MHz}}
\usepackage{listings}
\usepackage{color}
\usepackage{svg}

\usepackage{titlesec}

\begin{document}

\title{Atomic-optical interferometry in fractured loops: a general solution for Rydberg radio frequency receivers}

\author{Bartosz Kasza}
\email{b.kasza@cent.uw.edu.pl}
\affiliation{Centre for Quantum Optical Technologies, Centre of New Technologies, University of Warsaw, Banacha 2c, 02-097 Warsaw, Poland}
\affiliation{Faculty of Physics, University of Warsaw, Pasteura 5, 02-093 Warsaw, Poland}
\author{Sebastian Borówka}
\affiliation{Centre for Quantum Optical Technologies, Centre of New Technologies, University of Warsaw, Banacha 2c, 02-097 Warsaw, Poland}
\affiliation{Faculty of Physics, University of Warsaw, Pasteura 5, 02-093 Warsaw, Poland}
\author{Wojciech Wasilewski}
\affiliation{Centre for Quantum Optical Technologies, Centre of New Technologies, University of Warsaw, Banacha 2c, 02-097 Warsaw, Poland}
\affiliation{Faculty of Physics, University of Warsaw, Pasteura 5, 02-093 Warsaw, Poland}
\author{Michał Parniak}
\email{mparniak@fuw.edu.pl}
\affiliation{Centre for Quantum Optical Technologies, Centre of New Technologies, University of Warsaw, Banacha 2c, 02-097 Warsaw, Poland}
\affiliation{Faculty of Physics, University of Warsaw, Pasteura 5, 02-093 Warsaw, Poland}

\begin{abstract}
    The development of novel radio frequency atomic receivers brings attention to the theoretical description of atom-light interactions in sophisticated, multilevel schemes. Of special interest, are the schemes where several interaction paths interfere with each other, bringing about the phase-sensitive measurement of detected radio fields. In the theoretical modeling of those cases, the common assumptions are often insufficient to determine the boundary detection parameters, such as receiving bandwidth or saturation point, critical for practical considerations of atomic sensing technology. This evokes the resurfacing of a long-standing problem on how to describe an atom-light interaction in a fractured loop. In such a case, the quantum steady state is not achieved even with constant, continuous interactions. Here we propose a method for modeling of such a system, basing our approach on the Fourier expansion of a non-equilibrium steady state. The proposed solution is both numerically effective and able to predict edge cases, such as saturation. Furthermore, as an example, we employ this method to provide a complete description of a Rydberg superheterodyne receiver, obtaining the boundary parameters describing the operation of this atomic detector.
\end{abstract}

\maketitle

\section{Introduction}\label{sec:introduction}
\begin{figure}
\centering \includegraphics[width=0.8\columnwidth]{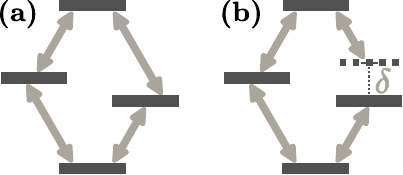} 
\caption{\textbf{(a)} Closed atomic loop. \textbf{(b)} Fractured atomic loop with a fracture $\delta$.}
\label{fig:simple_loop_compare}
\end{figure}
The modeling of continuous experiments in hot atomic vapors oftentimes relies on the presumption of a steady state, which is achieved by the atomic medium perpetually interacting with the E-M (electromagnetic) fields. This approach greatly simplifies the optical Bloch equations, removing the time-dependence, and has proven to be a very important tool in describing atomic-optical interactions semiclassically in many practical scenarios. However, there exists an inherent limitation to this approach, as the steady state cannot be achieved if the fields interacting with atoms form a fractured (as opposed to closed) interferometric loop, exemplified in the Figure \ref{fig:simple_loop_compare}, effectively beating via the atomic medium, such as when two separate fields are addressing the same transition. In particular, this lack of complete modeling approach becomes apparent in many cases of atom-based RF (radio frequency) receivers, where the presumption of steady state solution implies that their response to the detected fields cannot be modeled at non-zero bandwidth.

The consideration of loop-type interactions in atomic media is not a particularly recent topic. In the study of multi-level atoms, the loop-type schemes were investigated for 3-level \cite{Tsukada_1980} and 4-level loops \cite{Sharma_1984,Krinitzky_1986}. Such configurations were proposed as interfaces for phase sensitive measurements of E-M fields \cite{Buckle_1986} later expanded to more complex systems \cite{Kosachiov_1992} and particularly analyzed in a diamond-like configuration \cite{Morigi_2002,Kajari_Schr_der_2007}. The analysis of other effects in loop-type configurations included coherent population trapping \cite{Maichen_1996}, the investigation of EIT (electromagnetically induced transparency) phenomena in double $\Lambda$ setup, which showed that there is crucial impact of relative phases in closed loops \cite{Korsunsky_1999,Korsunsky_1999_2}, and the usage of the STIRAP (stimulated Raman adiabatic passage) method which lead to entanglement control of two-qubits systems \cite{Malinovsky_2004,Malinovsky_2004_2}. It is important to note that, in all of these considerations, only the cases of closed loops were fully addressed.

However, the topic has resurfaced in the context of Rydberg atomic RF receivers, where the atomic-optical interferometry brings a promise to enhance the measurements of MW (microwave) fields. A proposal for a Rydberg MW sensor utilizing a loop configuration was presented \cite{Shylla_2018}, and later experimental demonstrations appeared in different realizations \cite{Anderson_2022,Berweger_2023,Borowka_2024_2}. Employing various EIT effects in energy level loop configurations in this new kind of atomic-optical interferometer gave rise to a new method for phase-sensitive MW electrometry, yet its modeling is still limited to closed-loop schemes, which does not allow the theoretical considerations of the bandwidths of such receivers. Even though this discrepancy is most clearly demonstrated for diamond-like loop configurations, by extension it can be translated to the more prominent atomic mixers/superheterodyne receivers \cite{Simons_2019,Gordon_2019,Jing_2020}. These schemes can be treated analogously to fractured loops with the detuning present between LO (local oscillator) and signal field detected in a certain bandwidth.

Here we propose a general approach of solving fractured loop schemes based on Fourier decomposition of non-equilibrium steady state (NESS) \cite{Ikeda2020} applicable for any arbitrary loop fracture and energy level structure, so long as the driving fields can be uniquely mapped to energy levels transitions. Basing our approach on the density matrix formalism, we describe the necessary assumptions and transformations, finalizing with the Fourier mode decomposition, which is strictly connected with Floquet-Liouville supermatrix eigenvalue problem \cite{Ho_1985, Ho_1986}. As shown in multiphoton resonances case, the periodic, time-dependent equations are transformed into the time-independent matrix problem without using perturbation. This method is numerically effective and convenient to implement on any computational platform.

In particular, we use this method to simulate properties of a Rydberg RF superheterodyne receiver for typical experimental parameters. The considerations of loop fracture and signal field intensity allow for predictions of the superheterodyne receiver's bandwidth, saturation point and linearity -- the key performance parameters in RF detection. These parameters cannot be properly acquired with the small detuning and weak signal approximations, which are commonly used in the modeling of Rydberg superheterodyne receivers \cite{Jing_2020,Liu_2022,Ren_2024,Wu_2023}. We consider a model neglecting the Doppler effect, which needs to be introduced effectively as a correction in the further considerations. Therefore, we are effectively considering a single velocity class, which is often, practically, a reasonable approach. However, we do introduce the effects of transit-time broadening, which proved to be crucial in properly explaining the behavior of Rydberg atomic schemes \cite{Fan_2015}.

This article is organized as follows. In Section \ref{sec:Theoretical formulation}, we introduce the model, the Hamiltonian in the presence of fracture $\delta$, and decoherence. In Section \ref{sec:Fourier expansion of Non-Equlibirium Steady State}, we apply Fourier series expansion to NESS and derive a numerically efficient solution by separating the ground state repopulation. In Section \ref{sec:Example: Rydberg superheterodyne detection}, we investigate an example of the superheterodyne detection scheme, for which we perform numerical calculations and analysis. We draw the conclusion in Section \ref{sec:Summary and discussion} and discuss possible further development of our model. 

\section{Theoretical formulation}\label{sec:Theoretical formulation}
\subsection{Interaction picture for interfering fields}

The starting point for describing the interaction of the atom with light is to choose the energy eigenstates basis $\ket{i}$ of $n$ atomic states. Then the Hamiltonian in the Schrödinger picture $H^S$ reads:
\begin{equation}
    \begin{split}
    \bra{i} H^S \ket{i} &= \hbar \omega_{Ai}, \\
    \bra{j} H^{S} \ket{i} &=  \Omega_{ij} \exp({- i \omega_{Lij} t}) + 
    \Omega^*_{ij}  \exp({ i \omega_{Lij} t}).
    \end{split}
\end{equation}
Above $\hbar \omega_{Ai}$ is the energy of the $i$-th state, $\Omega_{ij} =  \vec d_{ij} \cdot \vec{\mathcal{E}}_{ij}/\hbar$ is Rabi frequency for transitions $i \rightarrow j$ with dipole moment $\vec d_{ij} = -e\bra{j} \vec{r} \ket{i}$, frequencies $\omega_{Lij}>0$, and electric field vectors $\vec{\mathcal{E}}$ giving complex amplitudes and polarizations. The total electric field equals $\vec E =\sum_{ij}\Re \vec{\mathcal E}_{ij} \exp(- i \omega_{Lij} t)$.

\paragraph{Interaction picture.}
From the constructed Hamiltonian $H^S$, the time dependence can often be removed. To do this, we separate the diagonal generator of simple evolution $H^S_0=\sum_i \hbar\omega_{Ri} \ket{i}\bra{i}$, where $\hbar\omega_{Ri}$ are the reference energies and move to interaction picture. To do so, we define operator $Q =  \exp({-i H^S_0 t/\hbar})$. We calculate the interaction Hamiltonian $H^I = Q^\dag (H^S - H^S_0) Q$. The resulting Hamiltonian has elements of the form:
\begin{eqnarray}
\bra{i} H^I \ket{i}  &=& \hbar (\omega_{Ai}-\omega_{Ri}), \\
\bra{j} H^I \ket{i} 
&=& \Omega_{ij}\exp({-i(\omega_{Lij} - (\omega_{Rj}-\omega_{Ri}})) t)  \nonumber \\ 
&&+ \Omega_{ij}^*\exp({i(\omega_{Lij} + (\omega_{Rj} - \omega_{Ri}))t}). 
\label{Eq:HIod=Oijexp+cc}
\end{eqnarray}
We choose reference energies $\hbar\omega_{Ri}$ so that they form a ladder in resonance with respective light beams $\omega_{Lij}  = \omega_{Rj} -\omega_{Ri}$. Then the first off-diagonal term is independent from time $t$. The second term is a fast oscillating term and is thus neglected in the rotating wave approximation (RWA). Finally:
\begin{equation}
\begin{split}
    \bra{i} H^I \ket{i}  &= - \hbar\Delta_i \quad -\Delta_i \equiv \omega_{Ai}-\omega_{Ri},\\
    \bra{j} H^I \ket{i}  &= \Omega_{ij}.
        \label{def: main_definitions_hamiltonian_elements}
\end{split}
\end{equation}

\begin{figure}
\centering \includegraphics[width=\columnwidth]{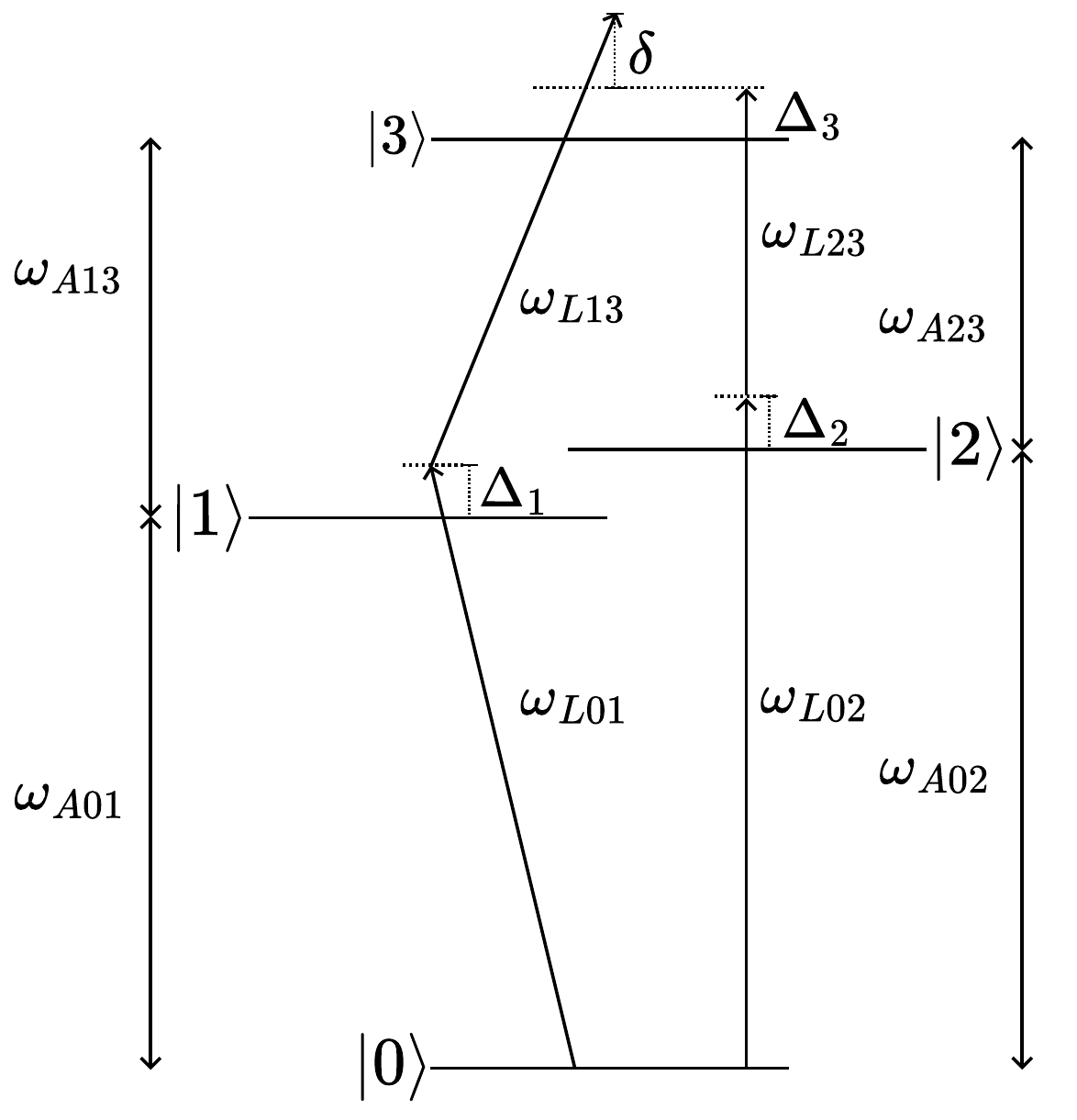} 
\caption{
4-level atomic fractured loop. It consists of four laser fields: probe laser of frequency $\omega_{L01}$ driving the transition $\ket{0} \rightarrow \ket{1}$, and laser fields $\omega_{L13}, \omega_{L02}, \omega_{L23}$ coupling the transitions $\ket{1} \rightarrow \ket{3}$, $\ket{0} \rightarrow \ket{2}$, $\ket{2} \rightarrow \ket{3}$ respectively. Dotted lines are determined by RWA reference energies $\omega_{Ri}$. The fracture occurs as left-hand side of atomic level ladder is mismatched by $\delta$ with the right-hand side. 
}
\label{fig:4p_loop}
\end{figure}

\paragraph{Imperfect RWA.}
We want to model an atomic-loop in the case when two fields meeting at an excited level have an energy mismatch, as presented in the Figure \ref{fig:4p_loop}. For this case, we tune the reference energy level to one of the fields, but the second one is mismatched by:
\begin{equation}
\delta_{ij}  \equiv \omega_{Lij}  - \omega_{Rj} + \omega_{Ri} \label{def: delta_definitions}.
\end{equation}
In this case, the first term in \eqref{Eq:HIod=Oijexp+cc} will be slowly oscillating with frequency $\delta_{ij}$, while the second will remain quickly oscillating with frequency $2\omega_{Lij}-\delta_{ij}$. Again, we neglect the fast oscillating term and get:
\begin{equation}
    \bra{j} H^I \ket{i}  = \Omega_{ij} \exp({-i \delta_{ij} t}), \quad
\end{equation}
In the above approach, we have assumed that each of the driving fields can be uniquely mapped to a single level transition. This requires in particular $\Delta_i, \delta_{ij}\ll \omega_{Lij}$. 

To set the stage for further analysis, let us point out the following useful decomposition: 
\begin{equation}
    H^I(t) = H_0 + H_+ e^{- i \delta_{ij} t} + H_- e^{i \delta_{ij} t}.
\label{eq:hamiltonian_separation}
\end{equation}
All of the further analysis will be done in interaction picture, we denote $H = H^I$.

In real physical systems, it is essential to introduce relaxation processes such as spontaneous emission. We use decay rate parameters $\gamma_{\alpha}^{\beta}$ for describing any incoherent transitions from the state $\ket{\beta}$ to the state $\ket{\alpha}$. The evolution of the density matrix $\rho$ of the system is given by the GKSL (Gorini-Kossakowski-Sudarshan-Lindblad) equation \cite{Gorini_1976,Lindblad_1976}:
\begin{equation}
    \begin{split}
       \dot{\rho}(t) &= -\frac{i}{\hbar}[H, \rho(t)] \\
       & + \sum_{\alpha, \beta} \gamma_{\alpha}^{\beta} \left(L^\beta_\alpha \rho(t) L^{\beta\dag}_\alpha - \frac{1}{2} \left\{L^{\beta\dag}_\alpha L^\beta_\alpha, \rho(t)\right\}\right), 
    \label{gksl}
    \end{split}    
\end{equation}
where $ L^\beta_\alpha = \ket{\alpha} \bra{\beta}$ describes transition $\ket{\beta}\rightarrow\ket{\alpha}$ and $\{{\cdot},{\cdot}\}$ is the anti-commutator. 

From this point, we also assume $\hbar = 1$. We can write the \eqref{gksl} equation as a Lindblad superoperator (Lindbladian) action on the density matrix:
\begin{equation}
    \dot \rho(t) = \superL \rho(t).
\end{equation}

We use a convention, commonly used in quantum information formalisms, where we define the non-Hermitian conditional Hamiltonian \cite{Fernengel2023, Porrati_1989} $H_c$ in such a way as to include the anticommutator from \eqref{gksl} to $H_c$, while the first term under the sum is separated as a repopulation superoperator $\superR$, transforming \eqref{gksl} into the form:
\begin{eqnarray}
        H_c &=& H - \frac{i}{2} \sum_{\beta} \Gamma_\beta \ket{\beta} \bra{\beta},  \quad \Gamma_\beta = \sum_{\alpha} \gamma_{\alpha}^{\beta},\\
        \superR \rho(t) &=& \sum_{\alpha, \beta} \gamma^\beta_\alpha
        \ket\alpha  
        \bra\beta \rho(t) \ket\beta 
        \bra\alpha.
        \label{eq:Rrho=sumgammarho}
\end{eqnarray}

The point of such a rewriting is based on the observation that the action of $H_{c}$ can be expressed as an evolution of the type $\exp(-i H_c t)$. Neglecting $\superR$, we get an equation describing the decay of excited states to the ground state.

In the numerical calculations, we use an explicit matrix forms of the Lindbladian:
\begin{equation}
    \superL_{ij}^{kl} = \left(H_c\right)_{ik} \mathds{1}_{jl} - \mathds{1}_{ik} \left(H_c^*\right)_{jl} + \superR_{ij}^{kl}. \label{full_lindbladian}
\end{equation}
where the repopulation operator $\superR$ is zero except terms $\superR^{\beta\beta}_{\alpha\alpha}=\gamma_\alpha^\beta$.

Significant numerical acceleration can be achieved by taking out a part $\rep$ of the repopulation $\superR$ responsible for repopulating the ground state $\ket{g}$. The part in question $\rep$ is required to satisfy $\rep \rho = \eta \cdot \ket{g}\bra{g}$ where we can find $\eta = \sum_\beta \gamma^\beta_g \bra{\beta} \rho \ket{\beta}$. This is confirmed by splitting \eqref{eq:Rrho=sumgammarho} into parts where the final state is the ground state $\alpha=g$. In other words the only nonzero terms in $\rep$ are:
\begin{equation}
    \rep_{gg}^{\beta\beta}=\gamma^\beta_g.
    \label{Eq:Gaagg=gammaag}
\end{equation}
The use of $\rep$ is illustrated via a method of calculating steady states presented in \ref{A:steadystate}.

\subsection{Evolution of $\rho$}

Let us now analyze the problem of a fractured loop, that is the set of fields interacting with atoms, which in the Dirac's picture cannot unambiguously form an interaction ladder with known detunings from resonant interactions. The simplest case of such a fractured loop is the case of two fields with different detuning interacting at the same atomic transition. From now on, the Hamiltonian $H_c$ is explicitly time-dependent $\bra{j}H_c(t)\ket{i} = \Omega_{ij}\exp(-i\delta t)$ for the field at the transition $\ket{i} \rightarrow \ket{j}$ with the fracture $\delta$. 
The unitary evolution of the state $\ket\psi$ from $t_0$ to $t$ would be described using the propagator $U(t, t_0)$,
so that $\ket{\psi(t+t_0)} = U(t + t_0) \ket{\psi(t_0)}$ and formally 
$U(t, t_0) = \mathcal{T} \exp(-i \int^t_{t_0} H_c(t') dt')$ 
where $\mathcal{T}$ is the time ordering operator. 
By analogy for our master equation with time-dependent Lindbladian $\dot{\rho}(t) = \superL(t) \rho(t)$,
a Floquet-Lindblad superoperator \cite{Chen_2024} can be expressed as $\superU(t, t_0) = \mathcal{T} \exp\left(\int_{t_0}^t \superL(t')dt'\right)$ and the density matrix evolution in time expressed by $\rho(t+t_0) = \superU(t, t_0) \rho(t_0)$.
For initial examinations of the evolution in our case, we approximate $\superU$ with a product of a sequence that is easily computed numerically. We use the first-order Suzuki-Trotter approximation:
\begin{equation}
\begin{gathered}
    \rho(t+t_0) \approx \prod_{s = {n-1}}^0 \exp\left(\superL (t_s) \Delta t\right) \rho(t_0),
\end{gathered}
\end{equation}
arriving at sequential approximation for each time step.

\begin{figure}
\centering \includegraphics[width=\columnwidth]{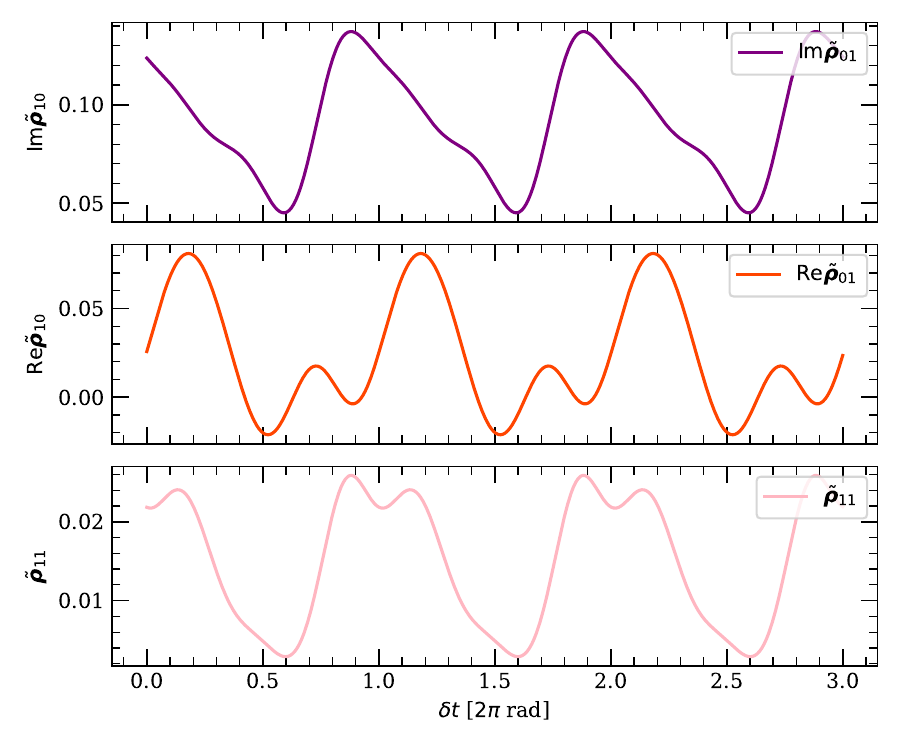} 
\caption{An exemplary evolution of NESS density matrix of 4-level atomic fractured loop presented on Figure \ref{fig:4p_loop}. We present three periods $T = 2\pi/\delta$ of NESS density matrix elements: $\cyclic_{01}, \cyclic_{11}$. The results are simulated for parameters: $\Omega_{01} = 1\unit$, $\Omega_{13} = 2\unit$, $\Omega_{12} = 5\unit$, $\Omega_{23} = 2\unit$, $\delta = 1\unit$, $\Gamma_1 = 6.1\unit$, $\Gamma_2 = \Gamma_3 = 0.1\unit$ assuming resonant fields: $\Delta_1 = \Delta_2 = \Delta_3 = 0$.}
\label{fig:evolution}
\end{figure}

\paragraph{NESS as eigenstate of one period of evolution.}
Knowing the sequential approximation formalism, we can determine the approximate eigenstate for the full sequence propagator. This is a NESS \cite{Ikeda2021, Chen_2024}, which we denote by $\cyclic$. As it is a periodic state $\cyclic(T+t_0) = \cyclic(t_0)$, we calculate it by finding the eigenstate of the propagator $\superU$ with an eigenvalue equal to~1:
\begin{equation}
\begin{gathered}
    \superU(T, 0) \cyclic (T)= \cyclic (T) = \cyclic (0).
\end{gathered}
\end{equation}
An example is plotted in the Figure \ref{fig:evolution}, where the $\cyclic (0) = \cyclic (T) = \cyclic (2T)$. 
The state evolves within a single rumbling sequence of frequency $\delta$.

\section{Fourier expansion of Non-Equilibrium Steady State}\label{sec:Fourier expansion of Non-Equlibirium Steady State}

Due to effective periodic driving condition and the resulting periodicity of the NESS, the density matrix can be decomposed into Fourier series in which only integer harmonics of $\delta$ appear: 
\begin{equation}
\cyclic(t) = \sum_{m=-\infty}^{\infty}\cyclic^{(m)} e^{i m \delta t}.
\end{equation}

A suitably decomposed Lindbladian can be obtained from the decomposed form of Hamiltonian \eqref{eq:hamiltonian_separation}:
\begin{equation}
    \superL(t) = \superL_0 + \superL_+ e^{- i \delta t} + \superL_- e^{i \delta t}.
\end{equation}

After a long-time interaction with the periodic field, we expect the harmonics to stabilize at a steady state values $\dot{\cyclic}^{(m)} = 0$.
Then the GKSL equation $\dot{\cyclic} = \superL(t) \cyclic $ is satisfied at each instance of time. 
On the left hand side, the derivative of the postulated form of the density matrix is: 
\begin{equation}
\dot{\cyclic}  = \sum_{m}(i m \delta \cyclic^{(m)} + \dot{\cyclic}^{(m)}) e^{i m \delta t},
\end{equation}
while on the right hand side, after collecting common Fourier exponents:\begin{align}
\superL \cyclic 
&= \sum_{m}(\superL_0 \cyclic^{(m)} + \superL_+ \cyclic^{(m+1)} + \superL_- \cyclic^{(m-1)} )e^{ i m \delta t}.
\end{align}
Comparing the coefficients side by side with the same time dependence $\exp(im\delta t)$, 
we come to the conclusion that for each mode $m$ in NESS, the following must be satisfied:
\begin{equation}
(\superL_0 - i m \delta) \cyclic^{(m)} + \superL_+ \cyclic^{(m+1)} + \superL_- \cyclic^{(m-1)} = 0.
\label{Eq:L0_d_rho}
\end{equation}
We have obtained a linear problem, which is referred to as Floquet-Liouville supermatrix (FLS) \cite{Ho_1985}. 

\paragraph{Difficulties with the numerical solution.}
The solution $\cyclic$ should be sought in the nullspace of the FLS \eqref{Eq:L0_d_rho}. 
However, this problem is numerically ill-posed, because in practice, 
we have to truncate the Fourier expansion to a finite number $m$ and thus the eigenvalues of the matrix may move away from zero. 

for larger matrix sizes

\paragraph{Separating out non-zero right-hand side.}
To circumvent the above difficulties we separate the average ground state repopulation (i.e. for $m=0$). 
We subtract $\rep$ \eqref{Eq:Gaagg=gammaag} from the Lindbladian and transfer to the right side of the equation for $m=0$: 
\begin{equation}
     (\superL_0 - \rep)\cyclic^{(0)} + \superL_+ \cyclic^{(1)} + \superL_- \cyclic^{(-1)} = -\rep \cyclic^{(0)}.
\end{equation}
Separately we use $\rep \cyclic^{(0)} = \eta \ket{g}\bra{g}$, 
where $\eta = \sum_\beta \gamma^\beta_g \bra{\beta}\cyclic_0\ket{\beta}$ becomes a normalization factor. We can solve for unnormalized density matrix $\bar \cyclic^{(m)}  = \cyclic^{(m)} / \eta$. 
The resulting equation is numerically efficient tridiagonal matrix linear problem:
\begin{widetext}
\begin{equation}
\begin{gathered}
     \begin{pmatrix} 
    \ddots & \superL_+ & 0 & 0 & 0 & 0 & 0 \\[0.5em]
    \superL_- &\superL_0 + 2\delta  & \superL_+ & 0 & 0 & 0 & 0\\[0.5em]
    0 & \superL_- &\superL_0 + \delta  & \superL_+ & 0  & 0 & 0\\[0.5em]
    0 & 0 & \superL_- & \superL_0 - \rep  & \superL_+ & 0 & 0 \\[0.5em] 
    0 & 0 & 0 & \superL_- & \superL_0 - \delta & \superL_+ & 0\\[0.5em]
    0 & 0 & 0 & 0 & \superL_- & \superL_0 - 2\delta & \superL_+ \\[0.5em]
    0 & 0 & 0 & 0 & 0 &  \superL_- & \ddots\\
    \end{pmatrix}
    \begin{pmatrix}
        \vdots\\[0.5em]
        \bar{\cyclic}^{(-2)} \\[0.5em]
        \bar{\cyclic}^{(-1)} \\[0.5em]
        \bar{\cyclic}^{(0)} \\[0.5em]
        \bar{\cyclic}^{(1)} \\[0.5em]
        \bar{\cyclic}^{(2)} \\[0.5em]
        \vdots
    \end{pmatrix}
    =
    \begin{pmatrix}
        \vdots \\[0.5em]
        0 \\[0.5em]
        0 \\[0.5em]
        \ket{g}\bra{g}\\[0.5em]
        0 \\[0.5em]
        0 \\[0.5em]
        \vdots
    \end{pmatrix}.
\end{gathered}
\label{Eq:tridiagonal matrix problem}
\end{equation}
\end{widetext}
We name FLS minus ground state repopulation for m=0 as FLSG. Solving the \eqref{Eq:tridiagonal matrix problem} linear equation, we are able to reconstruct the course of the NESS in time:
\begin{equation}
     \bar{\cyclic}(t) = \sum_m \bar{\cyclic}^{(m)} e^{i m \delta t}, \quad
     \cyclic(t) = \frac{1}{\Tr \bar{\cyclic}^{(0)}} \bar{\cyclic}(t).            
\end{equation}
Above we noted that $\Tr\cyclic=1$ and therefore $\eta=1/\Tr \bar{\cyclic}$.

In the numerical analysis, the range of $m$ has to be finite. In the further analysis, we always solve this problem for the symmetric range of modes, and consequentially for the central (with respect to $m = 0$) part of the FLSG, whose number of rows and columns considered is denoted by $N = 2|m_{max}|+1$, where $m_{max}$ is the number of outlying mode in the consideration. 
In numerical code, the problem \eqref{Eq:tridiagonal matrix problem} is solving linear equitation with three-index vectors 
--- one index for mode number $m$ and two indices for left and right hand states in the density matrix. The FLSG has therefore twice as many i.e. six indices. The vectors need to be reshaped to a single index columns and FLSG to a square matrix with two indices, then the standard linsolve is used and finally the resultant $\bar{\cyclic}$ is reshaped back to three-index form and normalized by the trace of the 0-th component.

\begin{figure}
\centering \includegraphics[width=\columnwidth]{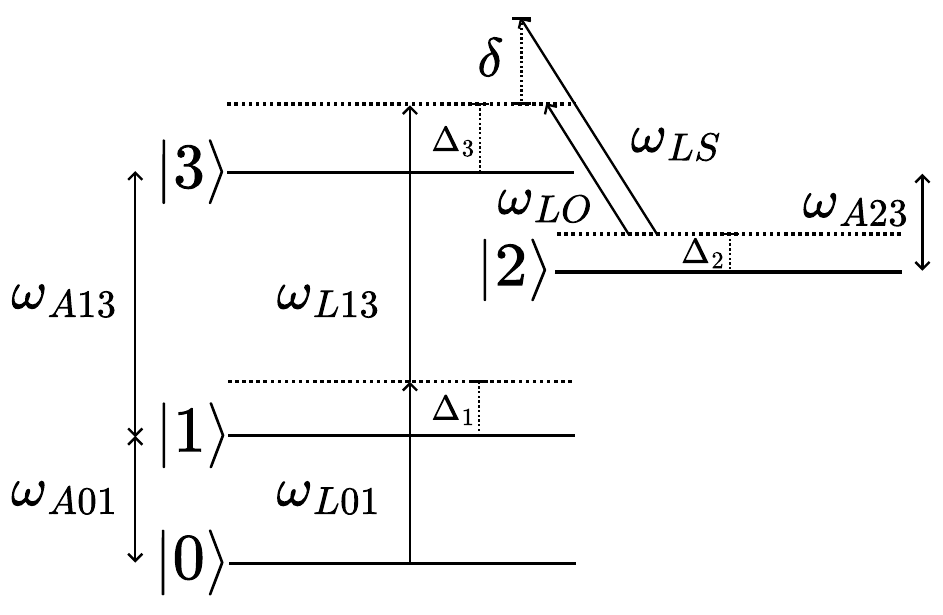} 
\caption{
    The energy level structure of a four-level Rydberg superheterodyne setup. It consists of two laser fields: 
    probe laser of frequency $\omega_{L01}$ driving the transition $\ket{0} \rightarrow \ket{1}$, and laser field $\omega_{L13}$ coupling the transition $\ket{1} \rightarrow \ket{2}$. A MW LO field drives the transition $\ket{2} \rightarrow \ket{3}$. Detunings from each levels respectively are $\Delta_{1}, \Delta_{2}, \Delta_{3}$. External MW signal field of an arbitrary detuning $\delta$ interacts with the system resulting in cyclic changes of probe transition that is measured and leads to the measurement of its spectral characteristics around the central frequency $\omega_{LS}$. 
}
\label{superheterodyne_scheme}
\end{figure}

\section{Example: Rydberg superheterodyne detection}\label{sec:Example: Rydberg superheterodyne detection}

\subsection{Hamiltonian}

Let us now consider the four-level atomic structure pictured schematically in the Figure \ref{superheterodyne_scheme}. The first two transitions are addressed by the optical laser fields, while the uppermost transition is in the MW domain. There, the LO MW field and the signal (LS) MW field interact in the realization of an atom-based mixer \cite{Simons_2019}, which we will further refer to as the superheterodyne detection scheme \cite{Jing_2020}.

To provide the description of the atom-light interactions in the superheterodyne scheme, we first consider the atom in the absence of the detected field LS. The atom is driven by two laser fields, probe and coupling, with frequencies $\omega_{L01}$, $\omega_{L13}$ respectively, addressing the transitions $\ket{0} \rightarrow \ket{1}$ and $\ket{1} \rightarrow \ket{3}$, and a LO MW field, driving the $\ket{2} \rightarrow \ket{3}$ transition. 
Typically the final level $\ket{2}$ is energetically below the intermediate level $\ket{3}$. 

Where we choose the rotating reference frequencies to be resonant with fields as depicted by dotted lines in the Figure \ref{superheterodyne_scheme}, 
which results in detunings from the energy levels:
\begin{align}
    \omega_{R_1} &= \omega_{L01} & \Delta_{1} &= -\omega_{A01} + \omega_{L01}, \\
    \omega_{R_2} &= \omega_{R_3} - \omega_{LO} & \Delta_{2} &= \omega_{A23} - \omega_{LO} + \Delta_3, \\
    \omega_{R_3} &= \omega_{R_1} + \omega_{L13} & \Delta_{3} &= -\omega_{A13} + \omega_{L13} + \Delta_1,
\end{align}
where $\omega_{Aij} \equiv \omega_{Aj} - \omega_{Ai}$ and we set $\omega_{A0} =0$. 
 
The addition of the LS field causes intensity rumbling of the MW field, which will be transferred onto absorption of the probe. The LS field drives the same $2\rightarrow 3$ transition as LO, but is detuned by $\delta$ from it. 
The MW field contributes to the $\bra{3}H\ket{2}=\bra{2}H\ket{3}^*$ term. 
The new term is time dependent with frequency $\delta$ because the RWA was tuned to LO field,
where from \eqref{def: delta_definitions} we find:
\begin{equation}
    \delta = \omega_{LS} - (\omega_{R3} - \omega_{R2}) = \omega_{LS} - \omega_{LO}.
\end{equation}
The Hamiltonian in the RWA is given by:
\begin{equation}
    H =  -\frac{1}{2}
    \begin{pmatrix}
        0 & \Omega_{01}^* & 0 & 0 \\[0.5em]
        \Omega_{01} & \Delta_{1} & 0 & \Omega_{L13}^*\\[0.5em]
        0 & 0 & \Delta_{2} & \Omega_{LO}^* + \Omega_{LS}^* e^{i \delta t}\\[0.5em]
        0 &  \Omega_{L13} &  \Omega_{LO} + \Omega_{LS} e^{- i\delta t} & \Delta_{3} \\
    \end{pmatrix}.
\end{equation}

This is a special case of the fractured loop, so we utilize our approach to find the dynamics of the NESS state, after driving with the effectively modulated field with beat note $\delta$. 

\subsection{Absorption and detection of its modulation}
We want to analyze the transfer of resulting field modulation to the coherence between states $\ket{0}$ and $\ket{1}$, which corresponds to the $\cyclic_{01}$ element of the periodic density matrix. In the typical detection scheme, where only the transmission through the atomic medium is measured, we can perform only the readout of the imaginary part of this element. Although the acquisition of the full form of the $\cyclic_{01}$ is possible with the interferometric techniques such as homodyne readout \cite{Kumar_2017,Ren_2024}, we will further focus on the $\Im \cyclic_{01}$.

In the experiment, we usually measure absorption $\propto \Im \cyclic_{01}(t)$, which can be written in a form of Fourier expansion:
\begin{equation}
    \begin{split}
    \Im \cyclic(t) 
    &= \frac{1}{2i}\left(\sum_m \cyclic^{(m)} e^{i m \delta t} - \sum_m \cyclic^{(m)*} e^{-i m \delta t} \right) \\
    &= \sum_m \underbrace{\frac{\cyclic^{(m)} - \cyclic^{(-m)*}}{2i}}_{\alpha^{(m)}}e^{i m \delta t}.     
    \end{split}
\end{equation}
Note that this sum is real and therefore its expansion coefficients are mutually conjugated $\alpha^{(m)}=\alpha^{(-m)*}$. We may therefore, use only positive sideband coefficients to reconstruct the result as:
\begin{equation}
    \Im \cyclic_{ij}(t) = \sum_{m\ge 0} \Re \left(\alpha^{(m)}_{ij} 
    e^{i m \delta t} \right).
\end{equation}

For now, we will focus on modulation transfer encoded in the $\alpha^{(\pm 1)}_{01}$ components of the above expansion, which add up to the modulation of the absorption of probe laser $\omega_{L01}$ at the same frequency $\delta$ as the detuning of the MW signal is applied at i.e. describe fundamental harmonic transfer. We follow analogous steps as in the experimental approach, so in order to find the modulation transfer, we need to apply demodulation $\mathcal{D}$ of order $\dbar$ for time $T$ to the absorption, so that:
\begin{equation}
    \alpha^{(\dbar)}_{01} = 
    \mathcal{D}^{(\dbar)}(\Im \cyclic_{01}) = \frac{1}{T} \int^{T}_0 \Im \cyclic_{01}(t) e^{- i \dbar t} dt \label{demod}
\end{equation}

Note that plugging the complex density matrix element $\cyclic_{01}$ into the demodulation operation would result in a corresponding Fourier coefficient, which in general is a different outcome than obtained in \eqref{demod} with the imaginary part. In this consideration, the modulus $|\alpha^{(1)}|$ gives the modulation transfer (the amplitude of probe field modulation), while the phase between MW modulation and probe modulation is given by $\arg \alpha^{(1)}$. 

\begin{figure}
\centering \includegraphics[width=\columnwidth]{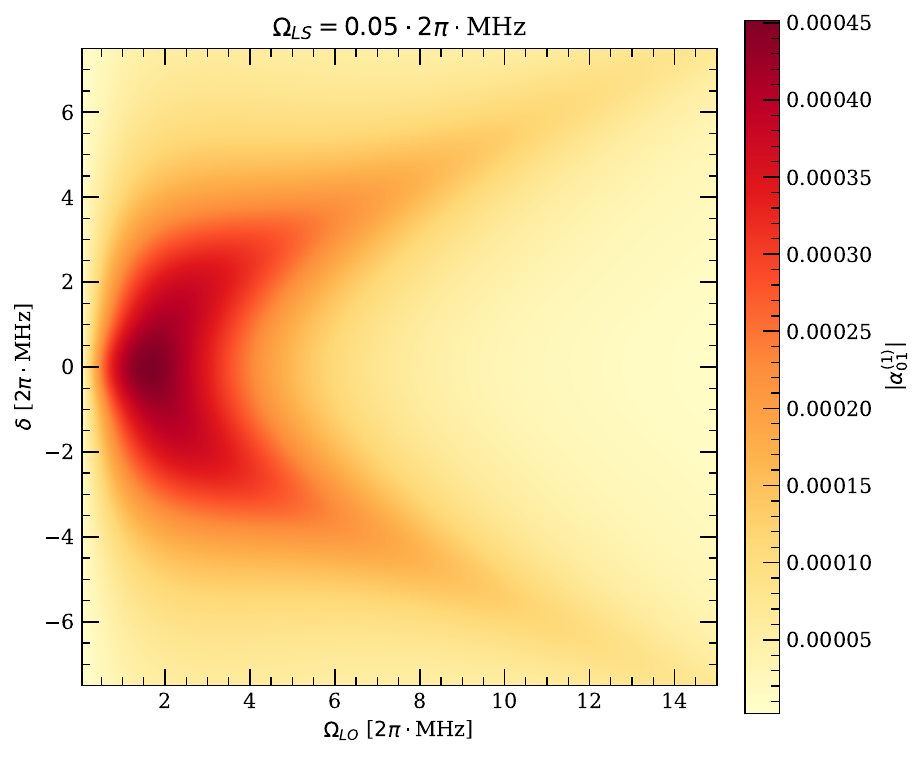} 
\caption{Modulation transfer $|\alpha_{01}^{(1)}|$ in superheterodyne setup as a function of fracture $\delta$ (vertical axis) and LO field $\Omega_{LO}$ (horizontal axis) for weak signal $\Omega_{LS}$. Enhancement of receiver's response is limited, as further increase of LO's Rabi frequency leads to A-T splitting. }
\label{sym_mode_LO_gt}
\end{figure}

\begin{figure*}
\centering \includegraphics[width=\textwidth]{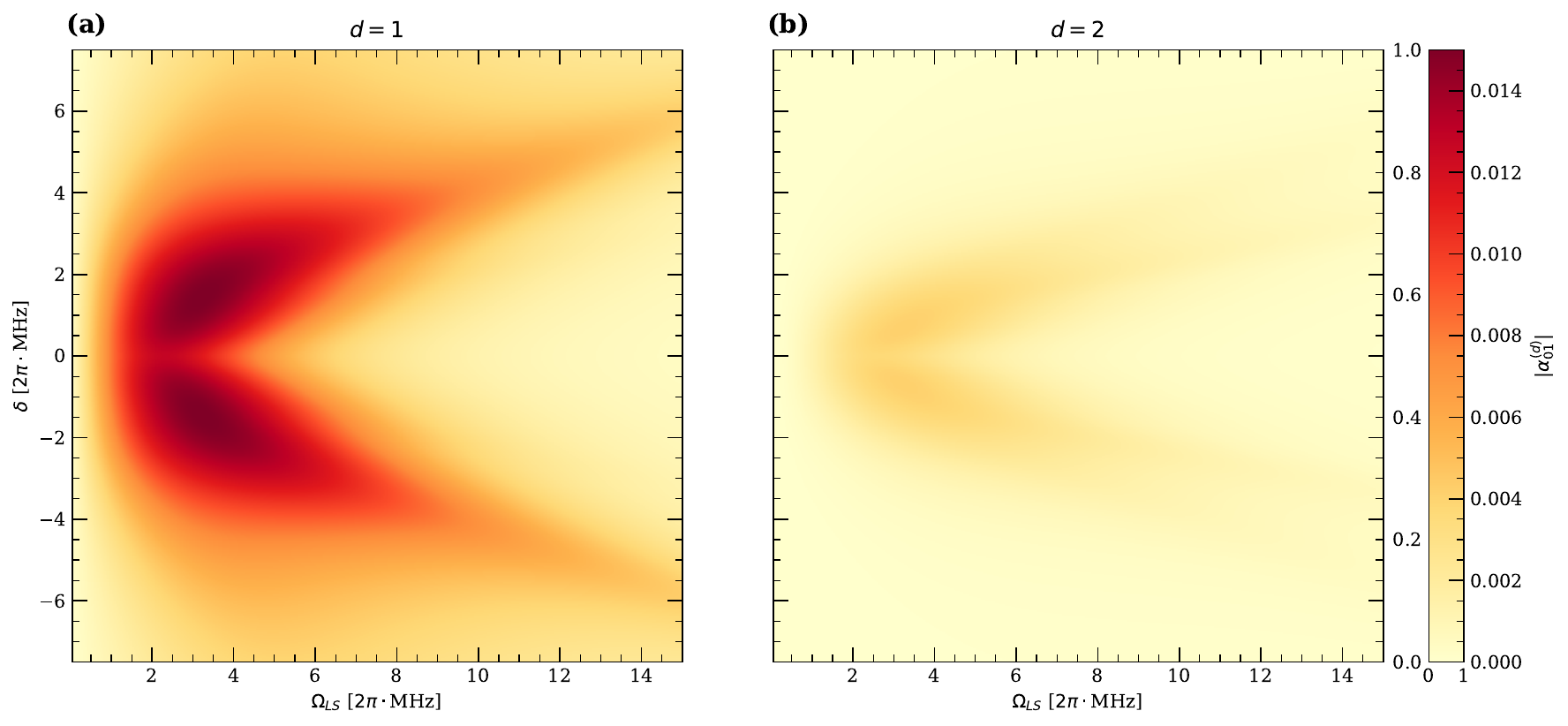} 
\caption{\textbf{(a)} Modulation transfer $|\alpha_{01}^{(1)}|$ in superheterodyne 4-level setup as function of fracture $\delta$ (vertical axis) and signal field $\Omega_{LS}$ (horizontal axis). Detected signal initially broadens with increase of $\Omega_{LS}$ until A-T splitting induces minimum in the center. \textbf{(b)} Absolute value of the second-order absorption demodulation $|\alpha_{01}^{(2)}|$ as function of fracture $\delta$ (vertical axis) and signal field $\Omega_{LS}$ (horizontal axis). As for $\dbar = 1$, it initially grows with $\Omega_{LS}$, but gets depleted with A-T splitting.}
\label{sym_mode_LS_gt}
\end{figure*}

\begin{figure}
\centering \includegraphics[width=\columnwidth]{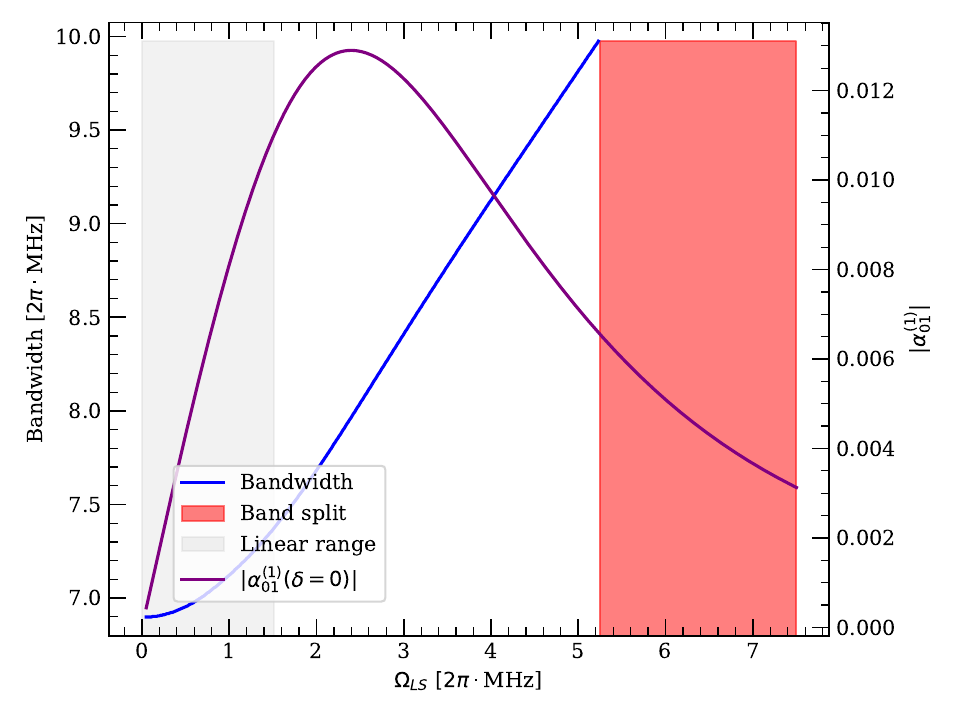} 
\caption{Bandwidth of the receiver as function of $\Omega_{LS}$ and modulation transfer $|\alpha_{01}^{(1)}|$, for which we choose the resonant case $\delta = 0$. The red area signifies the splitting of the band due to the minimum at the center getting smaller than half maximum due to the A-T splitting of energy levels induced by the strong signal field. The gray area signifies the linear response. The receiver saturates for $\Omega_{LS} = 2.4\unit$.}
\label{sym_bandwidth_LS_gt}
\end{figure}

\begin{figure}
\centering \includegraphics[width=\columnwidth]{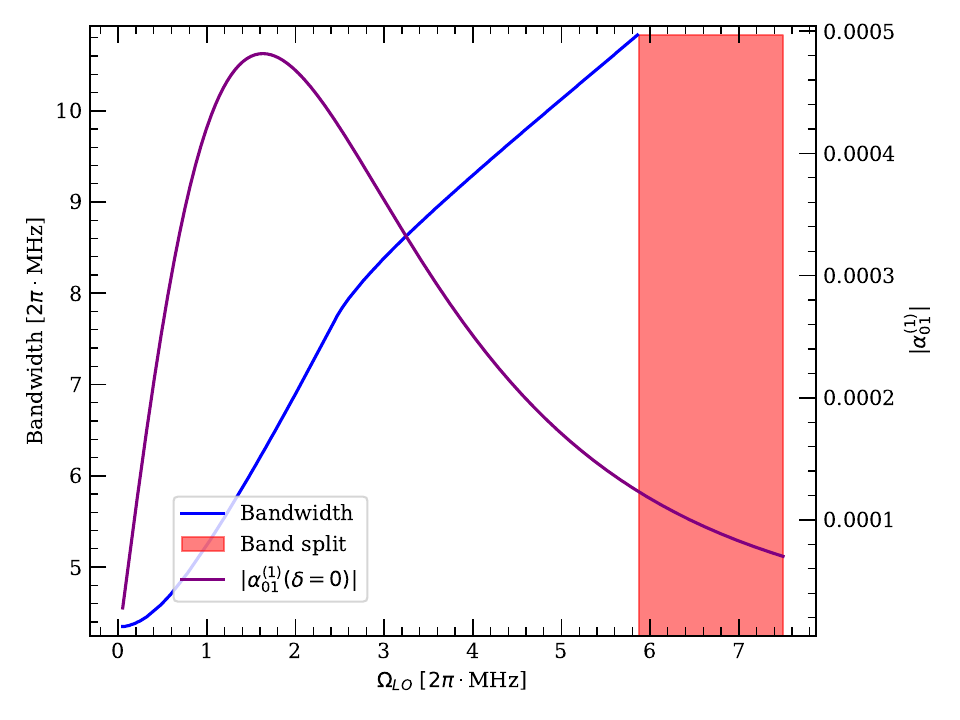} 
\caption{Bandwidth of the receiver as function of $\Omega_{LO}$ and modulation transfer $|\alpha_{01}^{(1)}|$, for which we choose the close resonant case $\delta = 0$. The red area signifies the splitting of the band due to the minimum at the center getting smaller than half maximum due to the A-T splitting of energy levels, induced by the strong LO field.}
\label{sym_bandwidth_LO_gt}
\end{figure}

\subsection{Results}

For numerical analysis, we consider the following values, similar to those typically employed in an experiment: probe $\Omega_{L01} = 1\unit$, coupling $\Omega_{L13}=5\unit$, for Rabi frequencies, $\Gamma_1 = 6.1\unit$ for decay. In the following consideration, Rydberg states' decay rates are limited by the transit-time broadening \cite{Fan_2015}, in the assumed case $\Gamma_{tr} = 0.8\unit$ ($\Gamma_0 = \Gamma_2 = \Gamma_3 = \Gamma_{tr})$, which also is added to the decoherence of the ground state. 
Moreover, let us consider the fully resonant case, typically optimal for the detection, 
where $\Delta_1 = \Delta_2 = \Delta_3 = 0$. 
For all simulations of modulation transfer at fundamental harmonic $\dbar = 1$, 
we limit the size of FLSG \eqref{Eq:tridiagonal matrix problem} to contain $|m|\le2$ Fourier harmonics; 
for calculating demodulation of order $\dbar = 2$, we use $|m|\le 3$. 
The analysis of precision depending on number of included Fourier modes is done in \ref{A: precision}.
An alternative example, applicable to cold atoms, where the natural decay rates dominate is presented in the \ref{A: cold atoms}. 

At first, we consider the dependence of the modulation transfer $|\alpha^{(1)}_{01}|$ on the Rabi frequency of the LO $\Omega_{LO}$, and the fracture $\delta$, which is equivalent to the detuning between the LO and the signal. Here, we assume weak signal field of $\Omega_{LS} = 0.05\unit$. The results are presented as a colormap in the Figure \ref{sym_mode_LO_gt}. The receiver's response is initially enhanced with increasing the LO field, but at one point, the energy levels become split due to the A-T (Autler-Townes) effect and the receiver's effectiveness is depleted. We find the optimal operating point at $\Omega_{LO} = 2\unit$ and apply this value in further considerations, resembling the procedure of experimental optimization of the receiver.

Now, we analyze sensing properties of the receiver with respect to signal field, Rabi frequency, and fracture $\delta$. Apart from the modulation transfer $|\alpha^{(1)}_{01}|$, we also investigate the absolute value of the second order demodulation $|\alpha^{(2)}_{01}|$. The results are presented as colormaps in the Figure \ref{sym_mode_LS_gt}. Analogously, as in LO dependency, we observe initial enhancement of modulation transfer that splits due to the A-T effect for stronger $\Omega_{LS}$. In the case of $\dbar = 2$, the A-T effect is visible and the importance of the second modes is notable, as at maximum point ($\Omega_{LS} = 3.3, \delta = -0.8$), the ratio $|\alpha^{(2)}_{01}|/|\alpha^{(1)}_{01}| = 30\%$, however, it is mostly negligible for weaker fields.

One of the most important parameters of the receiver is its bandwidth. It is a measure of the range of frequencies that a receiver can detect and process, and here it may be found by analyzing FWHM (full width at half maximum) of the modulation transfer with respect to the fracture $\delta$. In the Figure \ref{sym_bandwidth_LS_gt}, we plot the bandwidth of the receiver in relation to the signal field's strength. For reference, we also provide the dependence of the intensity of the demodulated mode for the center of the band to help with bandwidth-efficiency trade-off analysis. An increase in bandwidth due to A-T splitting can be observed for stronger fields, and at $5.2\unit$ the band becomes split, as a band gap appears in the center. Saturation of the receiver happens for $\Omega_{LS} = 2.4\unit$ as modulation transfer reaches its maximum value. The signal strength of $\Omega_{LS} = 1.5\unit$ marks the end of linear range, which we define as the point where the first derivative is smaller than half of its maximum value. 

Furthermore, we investigate the case of bandwidth dependency on the LO strength, which can be seen as a means to widen the band with the trade-off of efficiency. We again assume sensing of the weak signal field, $\Omega_{LS} = 0.05\unit$. The Figure \ref{sym_bandwidth_LO_gt} shows the bandwidth and the strength of modulation transfer interplay with one another and lets us choose optimal operating points for a desired application. Notably, this LO-facilitated tuning range is useful only up to $5.9\unit$ of LO Rabi frequency, whereupon the band becomes split.

\section{Summary and discussion}\label{sec:Summary and discussion}

Overall, we have presented a complete and convenient method of approximating NESS solutions for periodic conditions. Starting from utilizing rotating wave approximation, we arrived to a periodic Hamiltonian in the interaction picture. Followingly, we separated Lindblad superoperators and expanded NESS into Fourier coefficients to arrive at time-independent FLS equation. We optimized the numerical problem using the repopulation operator separation to get rid of the matrix kernel problem and arrived at a simple linearly solvable problem. Furthermore, we applied the described method to a practical example of Rydberg superheterodyne detection. Thanks to the versatility of the model and exact dependency on fracture $\delta$, we were able to efficiently find parameters of receivers, i.e.~dynamic ranges, saturation points, and bandwidths, which are crucial for modeling of such devices. These key parameters are of particular importance in light of the analysis of various tuning schemes \cite{Berweger_2023_2} and the increasingly frequent practical realizations of Rydberg-based detection, such as satellite \cite{Elgee_2023,Arumugam_2024} and WiFi \cite{Nowosielski_2024} signal sensing, as well as mmWave sources \cite{Legaie_2024,Bor_wka_2024}.

The proposed method has been exemplified on a Rydberg superheterodyne model, but in principle, it can be applied to the wider loop cases like the one presented in the Figure \ref{fig:evolution}. We hope that this method may bring the means to efficiently simulate the behavior of similar systems, offering enhancements to the MW detection \cite{Anderson_2022,Berweger_2023} and all-optical receiving \cite{Borowka_2024_2}. By extension, with the proper assumptions, the method can also bring some explanation to the proposed schemes for AM/FM receiving in Rydberg atoms \cite{Deb_2018,Holloway_2021,Anderson_2021,Bor_wka_2022}. Future work will include a full semi-algebraic treatment of the Doppler effect, which for the moment may be simply calculated numerically.

\section*{Data availability}
Data underlying the results presented in this paper are available in the Ref. \cite{DVN/TLKOJT_2024}.

\section*{Code availability}
The codes used for the numerical simulation are available from B.K.~upon request.

\begin{acknowledgments}
We thank Gabriel D.~Ko, Jan Nowosielski, Wiktor Krokosz and Mateusz Mazelanik for insightful discussions. The ,,Quantum Optical Technologies'' (FENG.02.01-IP.05-0017/23) project is carried out within the Measure 2.1 International Research Agendas programme of the Foundation for Polish Science co-financed by the European Union under the European Funds for Smart Economy 2021-2027 (FENG). This research was funded in whole or in part by National Science Centre, Poland grant no.~2021/43/D/ST2/03114.
\end{acknowledgments}

\appendix
\renewcommand{\thesection}{Appendix \Alph{section}}
\makeatletter
\renewcommand{\@seccntformat}[1]{\csname the#1\endcsname: }
\makeatother
\section{Steady state} \label{A:steadystate}

In the continuous experiment (i.e.~the experiment with constant, time-independent Hamiltonian), we often deal with the system some time after the conventional beginning of its evolution. After a sufficiently long time, we get a stationary state invariant to the evolution. We denote it by $\rhostd$. It is time independent $\dot{\rhostd} = 0$, that is, $\superL \rhostd  = 0$. According to this formula, in order to numerically find $\rhostd$, one would have to look for the null space of $\superL$.

To speed up numerical calculations, it is useful to transform the above problem into a linear equation with a non-zero right-hand side. To do this, we subtract the repopulation to the ground state $\rep$ from both sides of the equation for $\rhostd$, and use the property $\rep \rho = \eta \cdot \ket{g}\bra{g}$, where the number $\eta = \sum_\beta \gamma_g^\beta \bra{\beta}\rho\ket{\beta}$:
\setcounter{equation}{0}
\renewcommand{\theequation}{A\arabic{equation}}
\begin{equation}
    \begin{split}    
    (\superL - \rep)\rhostd & = -\rep\rhostd = \eta \cdot \ket{g}\bra{g} \\
    \implies \frac{\rhostd}{\eta} = \bar{\rhostd} & = (\superL - \rep)^{-1} (\ket{g}\bra{g}).\label{eq:separacja_repopulacji}
    \end{split}
\end{equation} 
In practice, solving \ref{eq:separacja_repopulacji} gives a non-normalized density matrix $\bar{\rhostd}$, which we divide by its trace $\rhostd = \bar\rhostd/\Tr\bar\rhostd$, recovering the sought steady state.

\begin{figure}
\centering \includegraphics[width=\columnwidth]{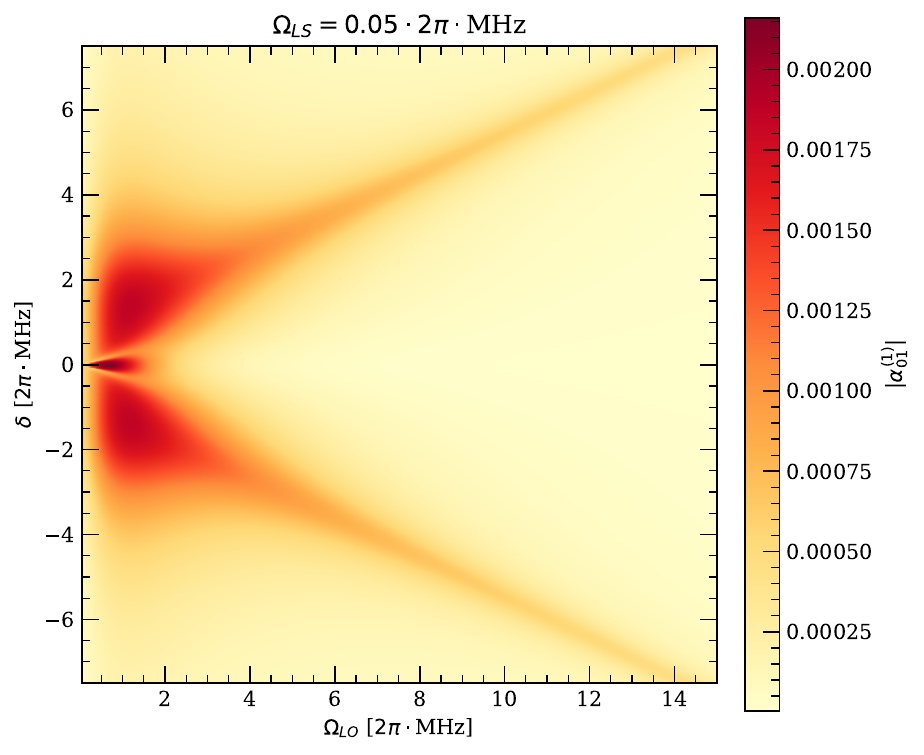} 
\caption{Modulation transfer $|\alpha_{01}^{(1)}|$ in superheterodyne setup as a function of fracture $\delta$ (vertical axis) and LO field $\Omega_{LO}$ (horizontal axis) for weak signal $\Omega_{LS}$ in cold atoms ($\Gamma_{tr} = 0$). The enhancement of receiver's response is limited, as further increase of the LO's Rabi frequency leads to A-T splitting, quicker than for the $\Gamma_{tr} > 0$ case. The maximum values area is narrow and limited for small fracture and weak LO. Additionally, there is a special regime, where for $\delta \approx 0$ and low $\Omega_{LO}$, the transfer of modulation is particularly efficient, which is not observed when transit-time broadening is considered.}
\label{sym_mode_LO_cold}
\end{figure}

\begin{figure*}
\centering \includegraphics[width=\textwidth]{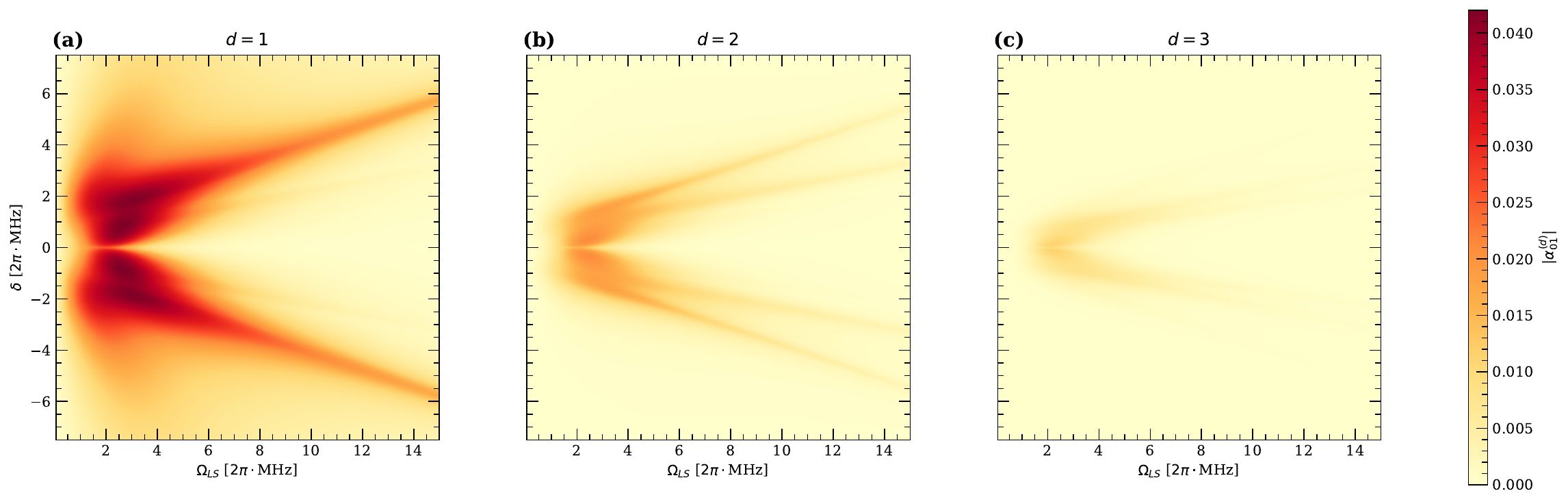} 
\caption{Comparison of absorption demodulation orders in superheterodyne 4-level setup, $|\alpha_{01}^{(1)}|$ as a function of fracture $\delta$ (vertical axis) and LO field strength $\Omega_{LO}$ (horizontal axis) for weak signal $\Omega_{LS}$ in cold atoms ($\Gamma_{tr} = 0$). \textbf{(a)} First order of demodulation. The A-T splitting is already significant for relatively weak signal. \textbf{(b)} Second order of demodulation. Two paths of A-T splitting are visible. \textbf{(c)} Third order of demodulation also undergoes A-T splitting, and overall, its impact is significantly weaker than those of the lower modes.}
\label{sym_mode_cold_1_2_3_LS}
\end{figure*} 

\section{Cold atoms} \label{A: cold atoms}
We provide an analogous analysis of the Rydberg superheterodyne detection scheme for cold, stationary atoms, that is for $\Gamma_{tr} = 0$, $\Gamma_{2} = 0.002\unit$, $\Gamma_{3} = 0.002\unit$ and setting all other values as before. On the plot, in the Figure \ref{sym_mode_LO_cold}, we see that for smaller $\Omega_{LO}$, we find $\delta \approx 0$ resonances; for stronger LO, we arrive, as described in the main text, at the linear regime of the modulation transfer. 

Similarly, for the considered LO strength of $\Omega_{LO} = 2\unit$, we analyze the transfer of modulation depending on the fracture $\delta$ and the signal strength $\Omega_{LS}$ and present the results in the Figure \ref{sym_mode_cold_1_2_3_LS}. Here, we extend our analysis to higher orders of demodulation, whose impact is more pronounced in the $\Gamma_{tr} = 0$ case. For demodulation order $\dbar = 1$, we set dimension of the FLSG \eqref{Eq:tridiagonal matrix problem} to $N = 5$; for $\dbar = 2$, we set $N = 7$; and for $\dbar = 3$, $N=9$. However, the impact from higher modes goes down significantly by the third order of demodulation.

\begin{figure}
\centering \includegraphics[width=\columnwidth]{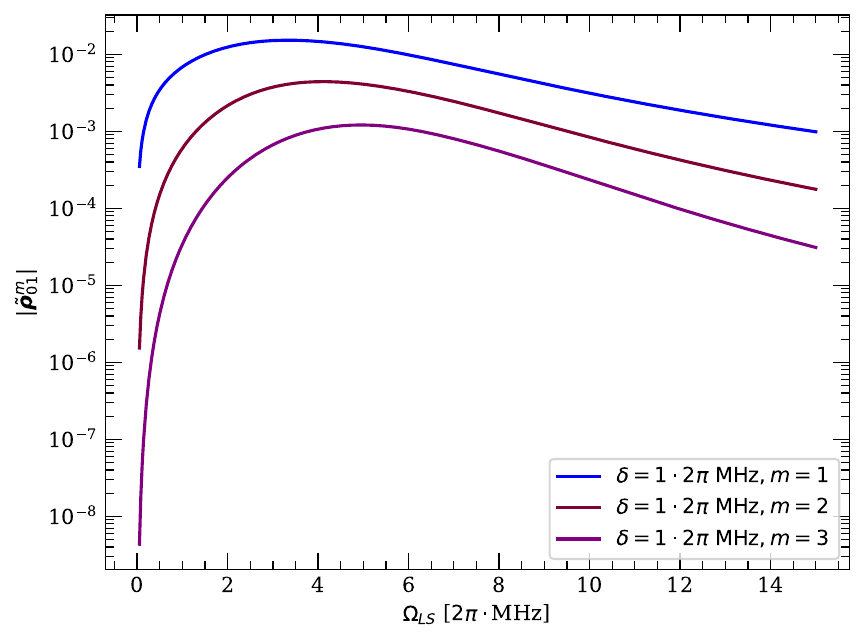} 
\caption{Comparison absolute values of Fourier coefficients $\rho_{10}^{(m)}$ as a function of signal $\Omega_{LS}$ for fracture $\delta = 1\unit$ and modes $m = 1, 2, 3$. For weak signals, one can neglect higher modes in the analysis and consider only the first mode.}
\label{mode_comparison_log}
\end{figure}

\begin{figure}
\centering \includegraphics[width=\columnwidth]{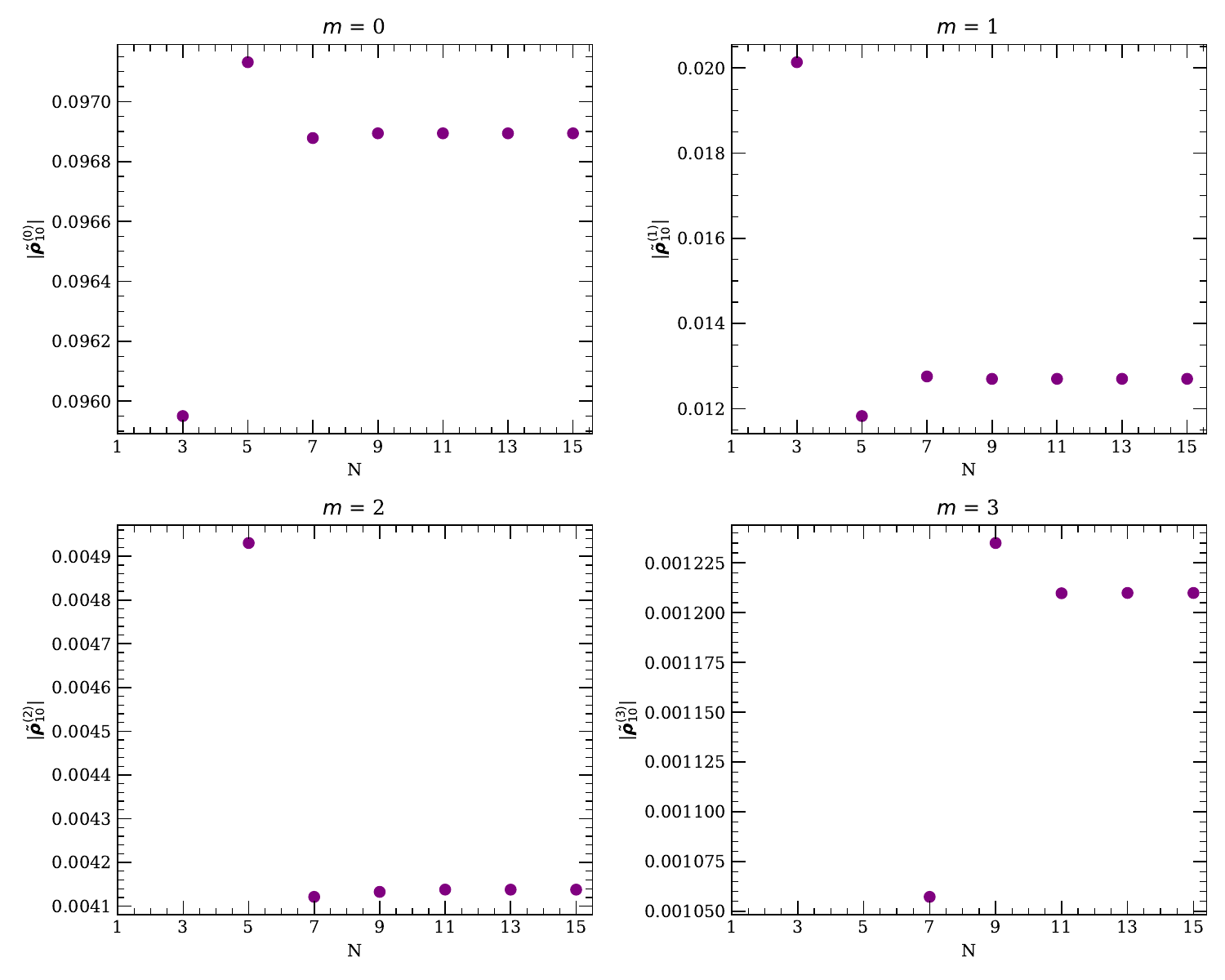} 
\caption{Comparison absolute values of Fourier coefficients $\rho_{10}^{(m)}$ for $ m = 0, 1,2, 3$ dependent on the dimension $N$. We chose to solve FLSG for the case of relatively high impact of $m=3$ setting $\Omega_{LS} = 4.9\unit$. The components quickly converge to correct values.}
\label{mode_comparison_lattice}
\end{figure} 

\section{Convergence and precision} \label{A: precision}

We investigate the importance of the $N$, understood as the number of rows and columns of the FLSG \eqref{Eq:tridiagonal matrix problem}, that needs to be considered to obtain a sufficiently accurate numerical solution. In particular, we expect that the solution converges numerically, that is the effects introduced by higher modes in the Fourier decomposition are consequently weaker. To illustrate this consideration, we set value of fracture $\delta = 1\unit$, so that the higher modes play a significant role in the evolution of atomic states in superheterodyne detection setup. We analyze the absolute value of Fourier coefficient $\cyclic^{(m)}_{01}$ dependency on signal $\Omega_{LS}$. From the Figure \ref{mode_comparison_log}, we choose a point, where third order $|\rho_{01}^{(3)}|$ is maximal, namely, we consider a strong signal field $\Omega_{LS} = 4.9\unit$ and fracture $\delta = 1\unit$. We plot the values of modulation transfer $|\alpha^{(\dbar)}_{01}|$ in this example for different ranges of modes spanning our FLS. The results are presented in the Figure \ref{mode_comparison_lattice}. As anticipated, the results converge for higher $N$ for all of the considered modes $m$. Furthermore, we see that it is often enough to set the $N$ one mode higher than the order of demodulation $\dbar$ that we want to analyze, so in the analyzed example $N_{\text{opt}} = 2(m + 1) + 1$ for our set of parameters. It should be noted, however, that in specific situations, where the nonlinear behavior of the system is more prominent, an even wider range $N$ may need to be considered.

\bibliographystyle{apsrev4-2}
\bibliography{bibliografia}
\end{document}